\documentstyle[aps,pre,amssymb]{revtex}
\input epsf.sty
\renewcommand{\theequation}{\arabic{section}.\arabic{equation}}

\begin{document}

\twocolumn[
\hsize\textwidth\columnwidth\hsize\csname@twocolumnfalse\endcsname

\draft

\title{ Self-Reduction Rate
of a Microtubule
}
\author{Takashi Hiramatsu$^1$, Tetsuo Matsui$^1$, and
Kazuhiko  Sakakibara$^2$}
\address{$^1$Department of Physics, Kinki University,
Higashi-Osaka, 577-8502 Japan
}
\address{$^2$Department of Physics, Nara National College of Technology,
Yamatokohriyama, 639-1080 Japan
}
\date{\today}

\maketitle

\begin{abstract}
We formulate and study a quantum field theory of a microtubule,
a basic element of living cells.
Following the quantum theory of consciousness  by
Hameroff and Penrose, we let the system to 
reduce to one of the classical states without measurement
if certain conditions are satisfied(self-reductions), 
and calculate the self-reduction time $\tau_N$
(the mean interval between two successive self-reductions)
of a cluster consisting of more than $N$ neighboring tubulins
(basic units composing a microtubule).
$\tau_N$ is interpreted there as an instance of the
stream of consciousness.
We analyze the dependence of $\tau_N$ upon $N$ and
the initial conditions, etc.
For relatively large electron hopping amplitude,
$\tau_N$ obeys a power law $\tau_N \sim N^b$,
which can be explained by
the percolation theory.
For sufficiently small values of the electron hopping amplitude,
$\tau_N$ obeys an exponential law, $\tau_N \sim \exp(c' N)$.
By using this law, we estimate the condition
for  $\tau_N $ to  take realistic values $\tau_N$
\raisebox{-0.5ex}{$\stackrel{>}{\sim}$}  $10^{-1}$ sec as
$N$ \raisebox{-0.5ex} {$\stackrel{>}{\sim}$} $1000$.

\end{abstract}


]


\section{Introduction}

Most approaches to understanding various characteristic
functions of the human brain in a framework of conventional
physics are classical. For example, the Hopfield model of neural
network\cite{hopfield}, a standard model of associative memory,
adopts an Ising spin variable ($\pm1$) to describe the state of
each neuron.
In the field of psychophysics, phenomenological relations
between psychological quantities (like the sense of pain) and
physical quantities
(like voltage) in classical physics are of main concern.
Although most researchers may agree that the human brain should be
described by quantum theory of electrons and atoms at the microscopic
level, there seems to be a prejudice that the functions of human
brain in our daily life are to be well understood, if possible,
by notion of classical physics without recourse to quantum physics.

The problem how to understand mind and consciousness
remains as a central and still open question
in the physics of the brain.
We don't know even whether the answer to this problem exists.
In such a situation, Hameroff and Penrose
\cite{hameroffandpenrose} proposed
a quantum theory of consciousness
according to a new quantum mechanics of microtubule.
Their idea is interesting because it suggests the
fundamental importance of quantum effects like quantum
measurements and collapses of wave function 
upon the functions of human brain.

\begin{figure}
  \begin{picture}(100,130)
\put(10,-10){\epsfxsize 80pt \epsfbox{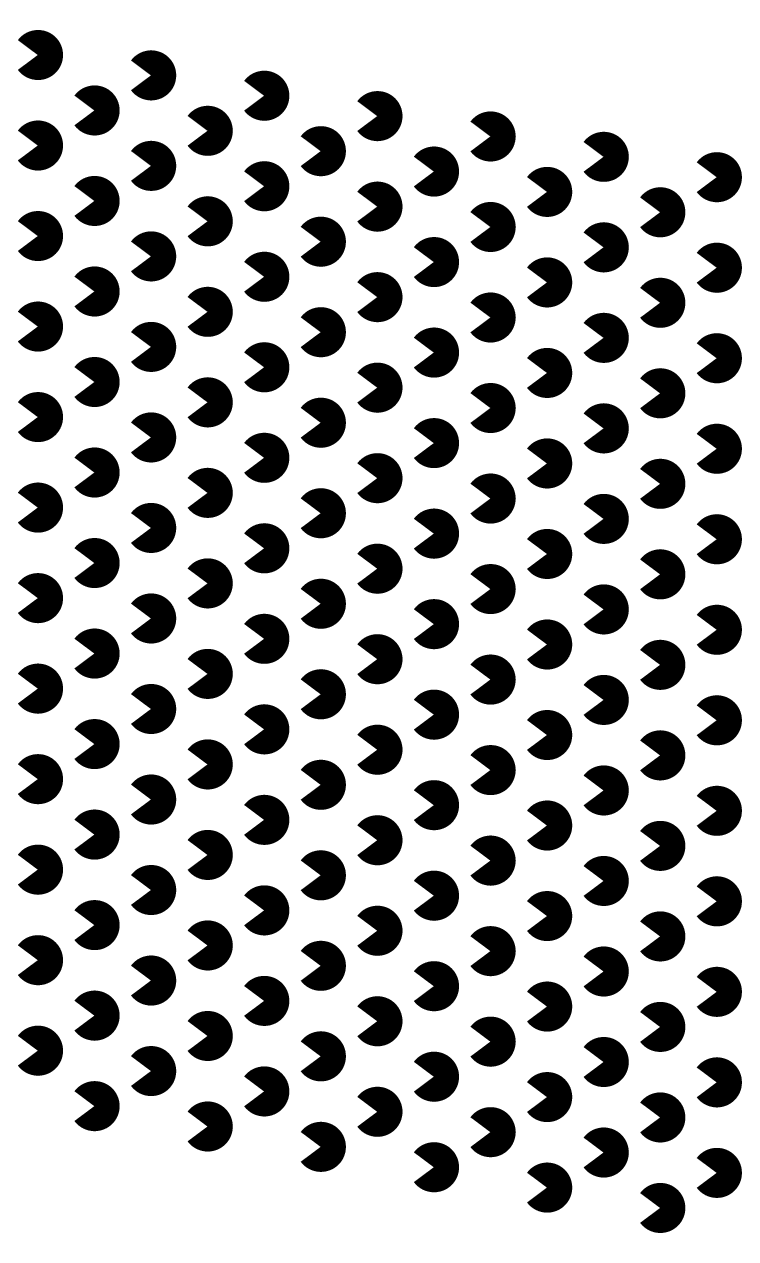}}
\hspace{3.3cm}(a) \hspace{-3.3cm}
\put(100,0){\epsfxsize 110pt \epsfbox{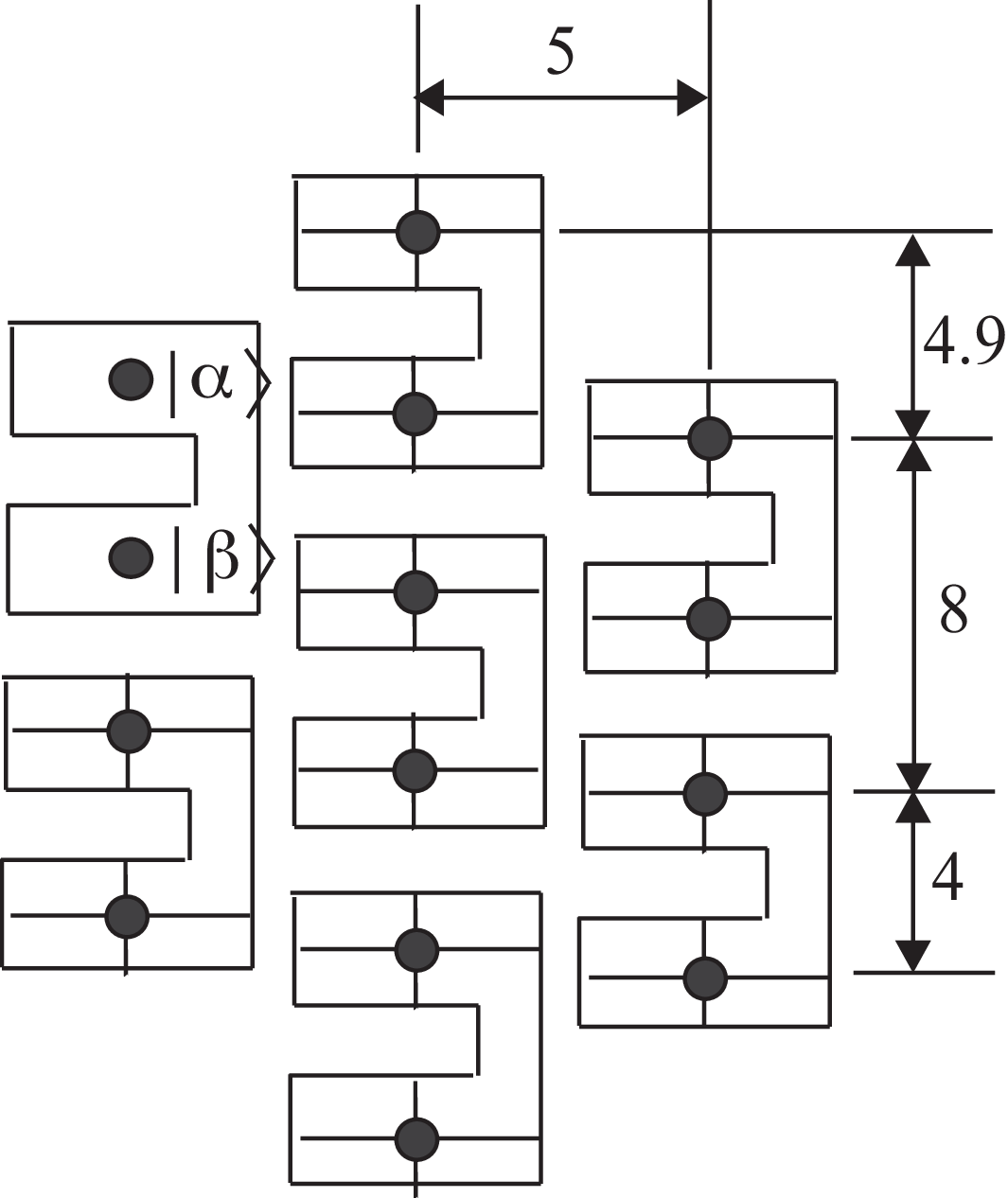}}
\hspace{6cm}(b)
  \end{picture}
  \vspace{0.5cm}
  \caption{(a) Development of a microtubule
  where each sector represents a tubulin and (b)
  Orientation of the six neighboring tubulins (from
Ref.[3]; lengths in unit of nm).
  Each filled circle represents a possible location of electron
  with corresponding quantum state $|\alpha \rangle$ (upper) or
  $|\beta \rangle$ (lower).@}
\label{fig1}
\end{figure}

Microtubules are basic building blocks of living cells
including neurons and axons, conveying mitochondria,
etc.
Each microtubule is a hollow cylinder with  a diameter 
$\sim 25$nm, the surface of which is
a 2D array of units called tubulins
(See Fig.\ref{fig1}a).
The array
consists of 13 columns, and its longitudinal size $L$
is not definite (typically $L=100 \sim 1000$).
Neighboring tubulins
in two adjacent columns is displaced with a fixed pitch,
so that tubulins form a triangular lattice.
A tubulin is a dielectric dimer consisting of
an $\alpha$ tubulin monomer and a $\beta$ tubulin monomer.

As a quantum model of a  tubulin it is often
treated\cite{hameroffandpenrose,tuszynski,rasmussen,penrose}
as each tubulin contains a single active(mobile) electron
and takes two independent states, $| \alpha \rangle$ and
$| \beta \rangle$, according to
the  "upper" or "lower" position
of the electron (See Fig.\ref{fig1}b).
Tuszy\'nski et al.\cite{tuszynski} set up and studied
some quantum models of a microtubule 
as a conventional quantum system, 
i.e., without self-reductions.
Also, Rasmussen et al.\cite{rasmussen} stuidied
signal propagations along a  microtubule by using
a network model of cellular automata.

Penrose \cite{penrose} argued that any quantum system is to be
affected through
couplings to quantum gravity in such a way that its wave function
should make
self-reductions (called orchestrated objective reductions)
according to certain rules that reflect the uncertainty
principle in quantum gravity.
In the theory of Hameroff and Penrose \cite{hameroffandpenrose},
a well developed
coherent (i.e., superposed) state of  each tubulin,
$ C_{\alpha} |\alpha \rangle + C_{\beta} |\beta \rangle $
with $|C_{\alpha}| \sim |C_{\beta}|$
has a too large uncertainty of the location of electron, and
may be  unstable if some other tubulins  are also
in such states at the same time, from the viewpoint
of uncertainty principle of quantum gravity. So each of
these tubulins  should make a self-reduction to be put back to
one of the eigenstates of electron position,
$|\alpha \rangle$ or $| \beta \rangle$.
They argue that a sequence
of such reductions 
works as a clock and forms a stream of consciousness.
The mean time between two successive reductions,
which we call the self-reduction time $\tau$,
is interpreted as each moment of consciousness.

Then quantitative estimation of $\tau$ would become a main concern.
However, Tegmark\cite{tegmark} posed a question that
a microtubule suffers from decoherence effects through
quantum interactions with its environment,
for instance ions and water molecules in the
surrounding cytoplasm,
and the decoherence time $\tau_{\rm dec}$
(the time scale to lose off-diagonal elements
of the reduced density matrix, hence to destroy long-range
quantum superpositions) may be much smaller
than the self-reduction time $\tau$. If this is true,
the scenario by Hameroff and Penrose loses its reality.
He   estimated
the decoherence time $\tau_{\rm dec}$ of an {\it entire}
microtubule at finite temperatures as
$\tau_{\rm dec}$ \raisebox{-0.5ex}{$\stackrel{<}{\sim}$}

\noindent
$ 10^{-13}$ sec and claimed that  microtubules
can be treated classically in the interesting
time scales  $10^{-1} \sim 10^{-2}$ sec
for a stream of consciousness,
and quantum effects play no significant role there.
Then Hagan et al.\cite{hagan}
responded to Tegmark by arguing in details that his estimate
is not definitive; $\tau_{\rm dec}$ may be much larger
and the quantum theory may be still relevant.

In this paper we leave this problem of estimating
the decoherence time as a still controversial and open problem,
and concentrate on the quantitative estimation of
the self-reduction time $\tau$ itself
of an isolated microtubule at zero temperature
without external disturbances, thus ignoring
decoherence effect by the environment.
An ensemble of microtubules
and/or the effects of temperature and
environment\cite{tegmark,hagan}
should be considered in the next step.
Explicitly, we set up a quantum theory
based on the classical model of Rasmussen et
al.\cite{rasmussen}  and calculate $\tau$ by
proposing some rules of self-reductions
based on the Hameroff-Penrose theory.
Such a study should certainly present some
important informations to scrutinize
the relevance of Hameroff-Penrose theory.
Since the typical time scale set by the Coulomb energy
at nano scales is $\sim 10^{-15}$ sec,
the central question is whether $\tau$ can take  values
of the order of $10^{-1} \sim 10^{-2}$ sec which
seems to be reasonable as the moment of consciousness.
We shall see that $\tau$ exhibits systematic behaviors
which may be typical for  models of self-reductions.

One may conceive reasons of self-reductions other than
Penrose's proposal based on quantum gravity \cite{penrose}.
Ghirard et al. \cite{drm} propose a modification
of Schr\"odinger equation to a nonlinear stochastic equation
to incorporate dynamical reductions. As yet another possibility,
(quasi)reductions may occur due to internal
quantum measurements of microtubule  by
surrounding environment in the brain,
or even the microtubule itself may ``measure" its part \cite{qe}.
Our rules of self-reductions are phenomenological
and may mimic these cases.

The rest of the paper is organized as follows:
In Sect.2 we set up a quantum field theory of  a microtubule.
It may be interpreted as a frustrated quantum spin model in
two dimensions. We propose to study its dynamics
by solving an approximate Schr\"odinger equation derived
by the variational method. In Sect.3 we calculate the self-reduction
time $\tau_N$ for various conditions.
We find that $\tau_N$ exhibits different behaviors
depending whether the electron hopping amplitude
is large or small. In Sect.4 we present conclusion.

\section{Model}
\setcounter{equation}{0}
\subsection{Hamiltonian}

The Hamiltonian
$H$
of a microtubule is given by
\begin{eqnarray}
H &=& -k \sum_{i}\sum_{\gamma=\alpha,\beta}
b_{i\bar{\gamma}}^{\dagger} b_{i\gamma}
+ \sum_{\langle i,j \rangle}\sum_{\gamma, \gamma'}V^{\gamma \gamma'}_{ij}
b_{i\gamma}^{\dagger} b_{i\gamma} b_{j\gamma'}^{\dagger} b_{j\gamma'},
\nonumber\\
V^{\gamma \gamma'}_{ij} &=& \frac{e^2}{\epsilon
R^{\gamma \gamma'}_{ij}},
\label{hamiltonian}
\end{eqnarray}
where $\bar{\alpha} \equiv \beta, \bar{\beta} \equiv \alpha$.
$b_{i\gamma}, b_{i\gamma}^{\dagger}\ (\gamma =\alpha, \beta)$
are the fermionic
annihilation and
creation operators of the electron in the $\gamma$ state
at the $i$-th tubulin, satisfying
\begin{eqnarray}
[b_{i\gamma},\ b_{i'\gamma'}]_+ = 0, \
[b_{i\gamma},\ b_{i'\gamma'}^{\dagger}]_+  =
\delta_{ii'}\delta_{\gamma \gamma'}.
\end{eqnarray}
$\epsilon$ is a dielectric constant of the microtubule.
The first term of $H$ represents the kinetic energy of the
electron
changing its state  $\alpha \leftrightarrow \beta$ {\it within}
the $i$-th tubulin with the hopping amplitude $k$.
The second term represents the Coulomb energy
$V^{\gamma \gamma'}_{ij}$
between two electrons at $(i\gamma)$ and
$(j\gamma')$ separated by the distance
$R^{\gamma \gamma'}_{ij}$, the value of which
is given in Ref.\cite{rasmussen}.
$\sum_{\langle i,j \rangle}$
implies the sum over all the neighboring pairs $i,j$
given in Fig.\ref{fig1}b (Each tubulin has six neighbors).
The number of electrons in each tubulin is a constant of motion,
\begin{eqnarray}
N_i \equiv \sum_{\gamma} b_{i\gamma}^{\dagger}b_{i\gamma},\
[H, N_i] = 0.
\end{eqnarray}
In actual calculations, we start and stay in the subspace
$N_i =  1$.

This Hamiltonian may be expressed in terms of the $s=1/2$
SU(2) quantum spin operator $\vec{S}_i \equiv (\hbar/2)
( b_{i\alpha}^{\dagger}, b_{i\beta}^{\dagger}  )
\vec{\sigma} \left( b_{i\alpha}, b_{i\beta}  \right)^t$
($\vec{\sigma}$ are Pauli matrices) as
\begin{eqnarray}
H &=& -\sum_{\langle i,j \rangle} J_{ij}S_{iz}S_{jz} -
\sum_{i}\vec{B} \vec{S}_{i} + {\rm const}, 
\label{heisenbergmodel}
\end{eqnarray}
with
\begin{eqnarray}
J_{ij}&=&-\frac{1}{\hbar^{2}}
\left( V^{\alpha \alpha}_{ij} - V^{\alpha \beta}_{ij}
- V^{\beta \alpha}_{ij} + V^{\beta \beta}_{ij} \right),
\nonumber\\
B_x  &=&  \frac{2k}{\hbar}, B_y= 0,\nonumber\\
B_z &=&-\frac{1}{2\hbar}  \sum_{j({\rm NN\ to\ } i)}
( V^{\alpha \alpha}_{ij} +  V^{\alpha \beta}_{ij}
-  V^{\beta \alpha}_{ij} -  V^{\beta \beta}_{ij} )=0.
\end{eqnarray}
The values of $J_{ij}$ are
\begin{eqnarray}
&&J_{\rm North}=J_{\rm South}=0.0833 \hbar^{-2}V_0,\nonumber\\
&&J_{\rm SE}=J_{\rm NW}=0.0091 \hbar^{-2}V_0,\nonumber\\
&&J_{\rm NE}=J_{\rm SW}=-0.0280 \hbar^{-2}V_0,\nonumber\\
&&V_0\equiv \frac{1}{\epsilon}\cdot \frac{e^2}{  1 {\rm nm}}=
\frac{1}{\epsilon}\cdot 2.31\times10^{-19}{\rm Joule},
\end{eqnarray}
where $V_0$ is the Coulomb energy of a pair of electrons
in the microtubule separated by 1 nm.
So the system may be viewed as a
frustrated spin model  in an external magnetic
field $\vec{B}$.
At $k=0$, the system involves
only $S_z$,
and its ground state is found to be the ``stripe state"
in which  $S_z$ are aligned to
$\pm \hbar/2$
along each column  with the alternative signs,
except for a pair of two NN columns (1st and 13th, say)
with the same signs.
The degeneracy is $13\times 2$\cite{wannier}.
As $k$ increases, the $S_x$
component develops, and  at $k\rightarrow \infty$,
all $\vec{S}_i$ align to $(\hbar/2)(1,0,0)$.

\subsection{Time evolution in variational method}
\setcounter{equation}{0}

In the time interval between two successive self-reductions,
the state vector $| \psi(t) \rangle$ of a microtubule at time $t$
evolves according to the Schr\"odinger equation,
\begin{eqnarray}
i \hbar \frac{d}{dt} | \psi(t) \rangle &=& H | \psi(t) \rangle.
\label{schroedinger}
\end{eqnarray}
If the microtubule contains $V$ tubulins,
then $| \psi(t) \rangle$ is $2^V$
dimensional, and for $V$ \raisebox{-0.5ex}{$\stackrel{>}{\sim}$} $13
\times 100$, the precise evaluation of
$| \psi(t) \rangle$ is
beyond the ability of our computers.
Thus, we evaluate $|\psi(t) \rangle$ approximately
by the variational method.
Explicitly, we choose the variational state
$|\psi_v(t) \rangle$ in the factorized form,
\begin{eqnarray}
&&| \psi_v(t)\rangle  =
\prod_i\ \left[\ {{C_{i\alpha}(t)\ |i \alpha \rangle
+ C_{i\beta}(t)\ |i \beta \rangle}_{}}^{}\ \right], \nonumber\\
&&| C_{i\alpha}(t) |^2 + | C_{i\beta}(t) |^2 =1,
\end{eqnarray}
where
\begin{eqnarray}
|i \gamma \rangle \equiv b^{\dagger}_{i \gamma} | 0 \rangle_i,\
b_{i\gamma} | 0 \rangle_i = 0,
\end{eqnarray}
is the state of the $i$-th tubulin in which
the electron is in the $\gamma$ state.
By minimizing the action,
\begin{eqnarray}
S = \int_{t_1}^{t_2} dt\ \langle \psi_v(t) | \left(
i\hbar \frac{d}{dt} -  H \right)\ | \psi_v(t) \rangle,
\end{eqnarray}
we get the  following equations of motion
for $2V$ complex coefficients $C_{i\gamma}$:
\begin{eqnarray}
i\hbar  \frac{d C_{i\gamma}}{dt} &=&
-k C_{i \bar{\gamma}} + J_{i\gamma} C_{i\gamma}, \nonumber\\
J_{i\gamma} &\equiv& \sum_{ j\; ({\rm NN\ to}\ i)}\left(
V_{ij}^{\gamma \alpha} | C_{j\alpha}|^2 +
V_{ij}^{\gamma \beta} | C_{j\beta}|^2  \right).
\label{EqOfM}
\end{eqnarray}
Eq.(\ref{EqOfM}) respects the unitarity,
\begin{eqnarray}
| C_{i\alpha}(t) |^2 + | C_{i\beta}(t) |^2 =1,
\end{eqnarray}
and keeps the average $E_v \equiv
\langle \psi_v(t) | H | \psi_v(t) \rangle$
a constant of motion.

We note that,
if  $\Delta V_{ij}^{\gamma} \equiv V_{ij}^{\gamma\alpha}
- V_{ij}^{\gamma\beta}$
is negligible,
\begin{eqnarray}
J_{i\gamma}\simeq \frac{1}{2}\sum_{j}
(V_{ij}^{\gamma \alpha}+V_{ij}^{\gamma\beta})
\equiv \tilde{J}
\end{eqnarray}
becomes independent of neighboring $C_{j\gamma}$'s, so
Eq.(\ref{EqOfM}) are decoupled to give the mean-field
solution (MFS),
\begin{eqnarray}
C_{i\alpha(\beta)}&=& C_{i+}e^{-i\Omega_+ t}+(-) C_{i-}
e^{-i\Omega_- t},\nonumber\\
C_{i\pm}&\equiv& \frac{1}{2}\left[C_{i\alpha}(0)\pm
C_{i\beta}(0)\right],\
\Omega_{\pm}\equiv\tilde{J}\mp k.
\label{mf}
\end{eqnarray}
The actual value of $|\Delta V_{ij}^{\gamma}|/
(V_{ij}^{\gamma \alpha}+V_{ij}^{\gamma \beta})$
averaged over $j$ is  0.19,
so the couplings between neighboring tubulins are intermediate.

We  solve the differential equations (\ref{EqOfM})
numerically  by the Runge-Kutta method with
discrete time step $\Delta t$.
As the initial condition
$C_{i\gamma}(0)$ at $t = 0$, we consider the following four cases:

\begin{eqnarray}
&{\rm (i)}\ &{\rm  US(Uniform\ Start);\ }  C_{i\alpha}(0) = 1,
C_{i\beta} = 0\ {\rm for\ all\ }i,
\nonumber\\
&{\rm (ii)}\ &{\rm SS(Stripe\  Start);\ }   C_{i\alpha}(0) = 1,0\
{\rm for\ alternative\ }\nonumber\\
&& {\rm columns\ as\ explained},
\nonumber\\
&{\rm (iii)}\ &{\rm RIS(Random\ Ising\ Start);\ }
C_{i\gamma}(0)=1,0, {\rm randomly},
\nonumber\\
&{\rm (iv)}\ &{\rm RCS(Random\ complex\ Start);\ }
C_{i\gamma}(0) \in {C}\ {\rm randomly}.\nonumber
\end{eqnarray}

\subsection{Rule of self-reductions}

For the rule of self-reductions, we follow the idea
of Hameroff and Penrose \cite{hameroffandpenrose}.
We first choose a parameter $P_0$ and
prepare the ``reset zone" of $C_{i\alpha}(t)$,
\begin{eqnarray}
P_0 \leq |C_{i\alpha}(t)|^2 \leq P_1(\equiv 1-P_0).
\end{eqnarray}
When $C_{i\alpha}(t)$ is in this zone,
we judge that the state $|i\alpha \rangle$ is sufficiently
coherent  and may need a self reduction.
In the simulation we watch whether $|C_{i\alpha}(t)|^2$
falls into this zone at every time step.
The tubulins that are in the reset zone form
a cluster, a set of tubulins that are connected
each other as neighboring pairs.
If the size of cluster at $t = t_R$ is {\it same or
more than} the prefixed number $N$,
we let all tubulins within this cluster make self-reductions
at the next time step.
As the rule to determine $C_{i\gamma}(t_R+\Delta t)$,
we consider the following two cases:

\begin{eqnarray}
{\rm LR}:&&  {\rm Local\ reduction;\ Each\ tubulin\ reduces\ to}\
|i\alpha \rangle\ {\rm or}\  |i\beta \rangle\nonumber\\
&&  {\rm according\ to\
the\ probabilities}\ |C_{i\gamma}(t_R)|^2.
\nonumber\\
{\rm GR}:&&  {\rm Global\ reduction;\  All\ the\ tubulins\
in\ the\ cluster\ reduce}
\nonumber\\
&&  {\rm to\ the\ common\ state,}\
|\gamma \rangle,\ {\rm with\ the\ probabilities\ }\nonumber\\
&&  \langle |C_{i\gamma}(t_R)|^2 \rangle, \  {\rm
the\ averages\ over\ the\
cluster.}\nonumber
\end{eqnarray}

\newpage

\begin{figure}
  \begin{picture}(0,105)
    \put(-10,0){\epsfxsize 230pt
    \epsfbox{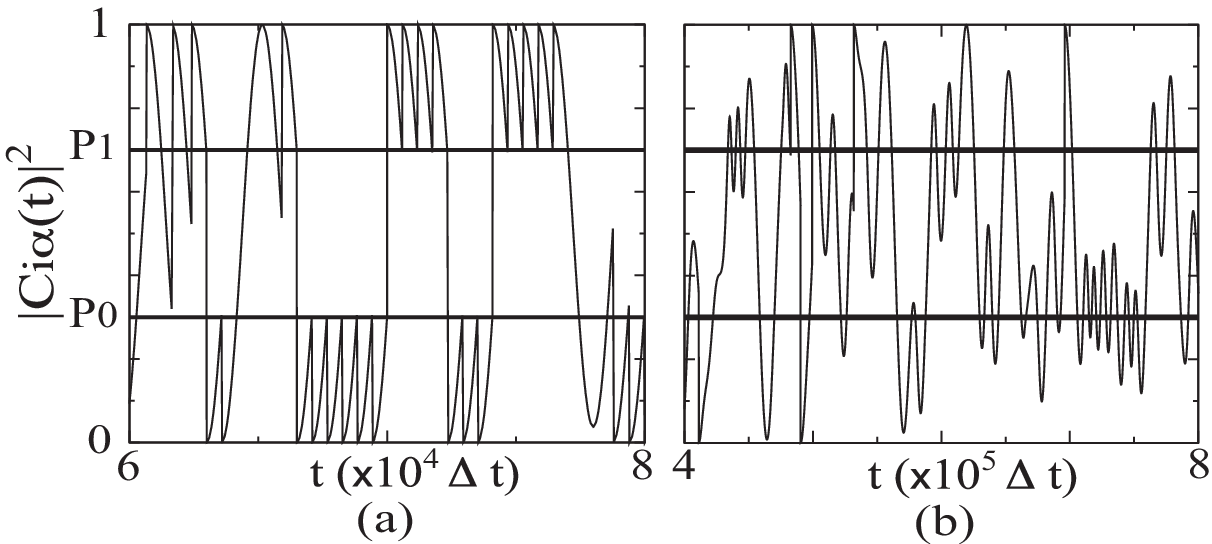}}
  \end{picture}
  \caption{Typical behavior of $|C_{i\alpha}|^2(t)$
in a run for RCS/LR, $L=100$.
(a) $k=0.1 V_0, N=100$ (b) $k=0.01 V_0, N=30$.
}
\label{fig2}
\end{figure}

In Fig.\ref{fig2} we show typical time dependence of
$|C_{i\alpha}(t)|^2$ for two values of $k$.
For larger $k$ the smooth motion between reductions
is rather regular,
reaching the endpoints 0 and 1 at every period.
For smaller $k$, the motion between reductions
is often pulled back before reaching the endpoints.
This difference is explained as the effect of potential term
in $H$ (See Sect.IIIa and IIIb).

\begin{figure}
  \begin{picture}(0,120)
    \put(0,10){\epsfxsize 220pt
    \epsfbox{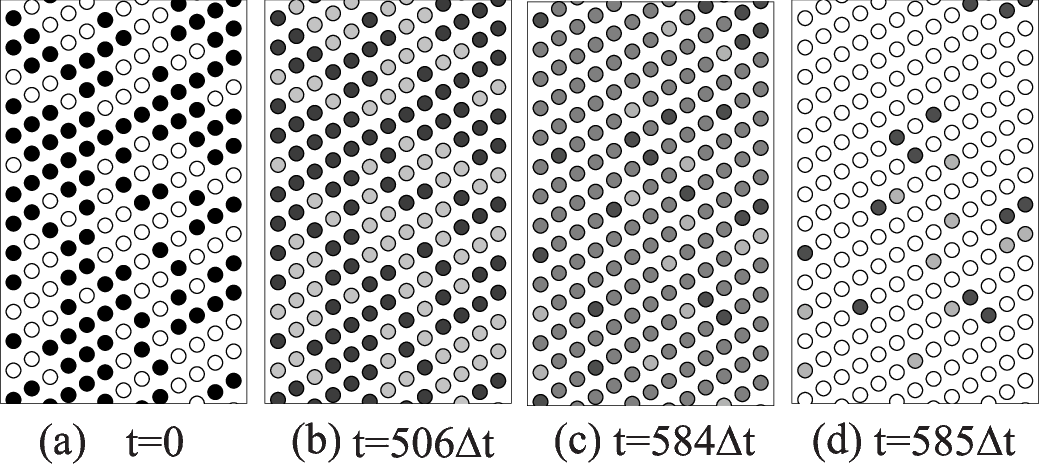}
}   \end{picture}
  \caption{Snapshots of time evolution of $|C_{i\alpha}|^2$
  in a part of a microtubule. Darkness
of each tubulin  represents  $|C_{i\alpha}|^2$ between
0(white) and 1 (black) in gray scales.
(a) RIS at $t=0$, (b) coherent evolution,
(c) sufficient number of tubulins are just in the reset zone
(a cluster is formed),
(d) most of tubulins are reset by the GR to $|\beta\rangle$ (white).
$k=0.1 V_0, N=200,L=100.$
}
\label{fig3}
\end{figure}

In Fig.\ref{fig3} we show typical snapshots of
$|C_{i\alpha}(t)|^2$  for a part of a microtubule
before and after a reduction.

To calculate the self-reduction time $\tau_N$ for the minimum
cluster size $N$, we make a sufficiently long run with the period
$t_{\rm total}$ for each $N$.
$\tau_N$  is then defined as

\begin{eqnarray}
\tau_N = \lim_{t_{\rm total}\rightarrow \infty}
\frac{t_{\rm total}}{M_N},
\end{eqnarray}
where $M_N$ is the total number  of all
the self-reductions within the run. So
$\tau_N$ is the average time interval between two successive
reductions. $\tau_N$ may be  expressed by
using the probability $p_N$ that
clusters of size equal to or larger
than $N$ appear as
\begin{eqnarray}
\tau_N &\propto&
\sum_{m=1}^\infty m(1-p_N)^{m-1} p_N = \frac{1}{p_N}.
\label{indep}
\end{eqnarray}

In the simulations we consider microtubules of the length $L$
up to $L =$ 2000 (16 $\mu$m)
with the open boundary  condition.
We choose

\noindent
\begin{eqnarray}
\Delta t&\equiv&  0.01 \times  t_0, \nonumber\\
t_0  &\equiv& \frac{\hbar}{V_0}=
\frac{\hbar \epsilon \times 1 {\rm nm}}{e^2}=
4.571 \times 10^{-16}\times \epsilon\ {\rm sec},
\end{eqnarray}
where $t_0$ is the time scale set by
the Coulomb energy $V_0$ at 1 nm.
We choose sufficiently large $t_{\rm total}$ up to
\begin{eqnarray}
t_{\rm total}=2\times 10^6 {\rm steps}\ \times \Delta t,
\end{eqnarray}
to obtain stable values of $\tau_N$.
The above $\Delta t $ is checked to be sufficiently small
for this $t_{\rm total}$.
To determine the clusters at each $t$,
we adopt the algorithm of Hoshen and Kopelman\cite{cluster}
developed in percolation theory\cite{percolation}.
In Table 1 we collect the parameters and their values
used in our simulations.

\vspace{0.5cm}
\hspace{-0.3cm}
\begin{tabular*}{8.8cm}{lll}\hline\hline
parameter & symbol & value
\\ \hline
hopping& $k$ &0.002 $\sim 2 (\times V_0)$ \\
amplitude & & \\ \hline
length & $L$ & 50 $\sim$ 2000 \\ \hline
& uniform  (US)& \\
initial & stripe  (SS)& All\\
condition  & random Ising (RIS)& \\
& random complex (RCS)&
\\
\hline
cluster size &$N$  & 5 $\sim$ 26000 \\
\hline
reset zone &$P_0$ & 0.01 $\sim$ 0.49 \\
\hline
reduction & global reduction (GR)& \\
condition & local reduction (LR) & both \\
\hline
elapsed time & $t_{\rm total}$ & $10^5 \sim 2 \times 10^6 (\times
t_0)$ \\
\hline\hline
\end{tabular*}
\begin{center}
Table 1. Parameters used in simulation.
\end{center}

\vspace{0.1cm}

We note that the Schr\"odinger equation (\ref{schroedinger})
is invariant under $t \rightarrow \lambda t$ and
$H  \rightarrow \lambda^{-1}H$. This is reflected
also in the equation of motion (\ref{EqOfM}).
Thus, $\tau$ is proportional to the dielectric constant
$\epsilon$ when $k$ is measured in unit of $V_0$.
Experimental measurements
of  $\epsilon$ are still not definitive but
predict values like
$\epsilon = 8.41$\cite{mershin}, $1 < \epsilon < 100$
\cite{minoura}, etc.
Also, if the charge of the
mobile object
which determines the tubulin state,
$|\alpha\rangle$ or $|\beta\rangle$, is
modified from $-e$ of a single electron to  $- ne$,
its effect is to be reflected by replacing
$\epsilon \rightarrow \epsilon/n^2$ in the formulae below.


\section{Self-Reduction Time}
\setcounter{equation}{0}

As we shall see, $\tau_N$ exhibits different $N$ dependences
for large $k$'s and small
$k$'s.  We  study the case of large $k$ first,
and then the case of small $k$.

\subsection{Large hopping amplitude $k$}

\begin{figure}
  \begin{picture}(30,160)
    \put(-60,10){\epsfxsize 500pt
    \epsfbox{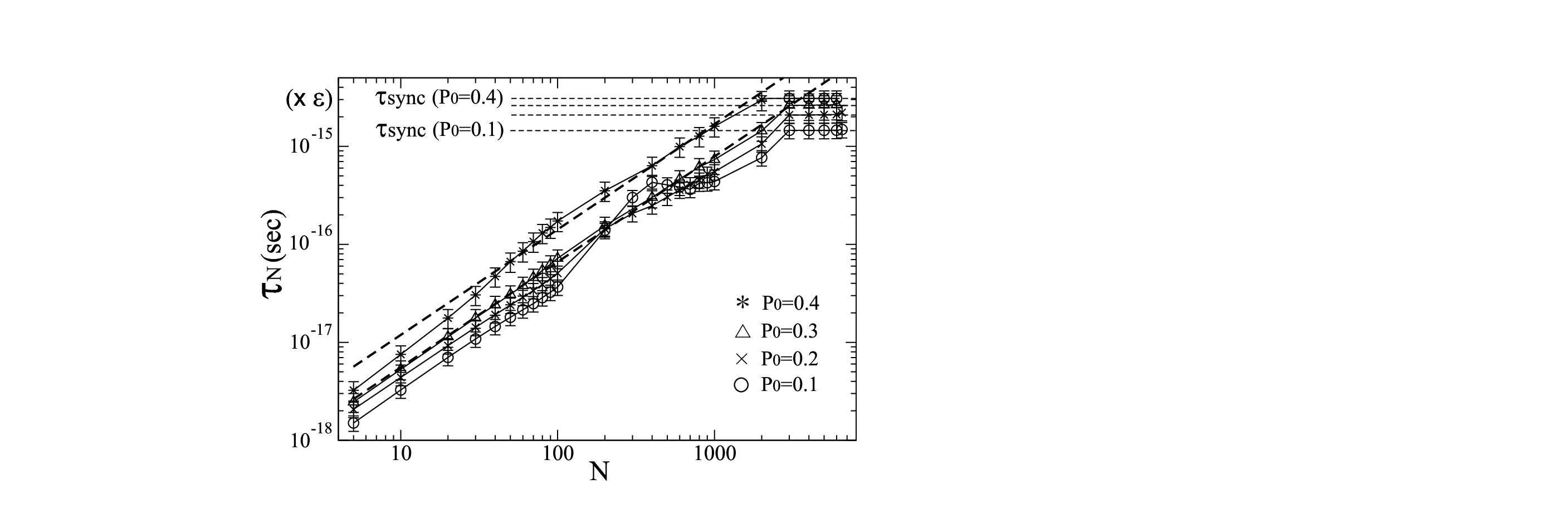}}
  \end{picture}
  \caption{$\tau_N$ vs. $N$
  for $k=0.1 V_0$ and $L=500$ with RIS/LR.
Errors are estimated by 20   independent runs.
The dashed straight lines show the power law of (\ref{power}).
The horizontal lines show the saturation
  value $\tau_{\rm sync}$ defined by Eq.(\ref{sync}).
}
\label{fig4}
\end{figure}

Let us first consider the case of large $k$.
In Fig.\ref{fig4} we plot $\tau_N$ vs.
$N$ for $k=0.1 V_0$ for several values of $P_0$.
For all the four initial conditions,
$\tau_N$ up to  $N \raisebox{-0.5ex}{$\stackrel{<}{\sim}$}
V/2$ can be fitted well by the power law,

\vspace{0.1cm}

\begin{eqnarray}
\tau_N &\simeq& a N^b \times t_0,\nonumber\\
a&\simeq&1.02\times10^{-3} t_0,\ b\simeq 1.08\ 
{\rm for}\ P_0=0.3.
\label{power}
\end{eqnarray}
Eq.(\ref{power}) may be explained by adopting the independent
oscillator
model where $|C_{i\alpha}(t)|^2$ is given by (\ref{mf}).
Then the probability $p$ that each $|C_{i\alpha}(t)|^2$ is in the
reset zone is given by
\begin{eqnarray}
p=1-\frac{2}{\pi}\cos^{-1}(1-2P_0).
\end{eqnarray}
For $P_0 \ge 0.146$, $p$ is {\it below} the
critical probability $p_c = 0.5$
of the triangular lattice above which percolation takes place
\cite{percolation}.
The scaling argument\cite{percolation}
predicts $p_N$ for $p < p_c$  as
\begin{eqnarray}
p_N \propto \sum_{s=N}^\infty s^{-\delta}f(z)
\propto
\left\{
\begin{array}{ll}
N^{-\delta+1}, &  (p_c-p)N^{\sigma} \raisebox{-0.5ex}{$\stackrel{<}{\sim}$} 
1,
\\
e^{-cN}, &  (p_c-p)N^{\sigma} \raisebox{-0.5ex}{$\stackrel{>}{\sim}$} 1,
\end{array}
\right.
\label{percolation}
\end{eqnarray}
where
$f(z), z\equiv (p_c-p) s^{\sigma},$
  is the scaling function
and the exponents are given by\cite{percolation}
\begin{eqnarray}
\delta=\frac{187}{91},\ \sigma=\frac{36}{91},\
c \propto (p_c -p)^{1/\sigma}.
\end{eqnarray}
By assuming
\begin{eqnarray}
(p_c-p)N^{\sigma} \raisebox{-0.5ex}{$\stackrel{<}{\sim}$} 1,
\end{eqnarray}
$\tau_N \propto
1/p_N \propto N^{\delta-1} $
giving us the form of
(\ref{power}) with
\begin{eqnarray}
b = \delta-1 = 1.055.
\end{eqnarray}

For $N \raisebox{-0.5ex}{$\stackrel{>}{\sim}$} V/2$,
only one large cluster is possible and all the tubulins in it
exhibit a synchronized (collective) motion as
\begin{eqnarray}
|C_{i\alpha}|^2 &\simeq& \rho_{\rm sync}(t) \equiv
\cos^2[\frac{k}{\hbar}(t-t_R)]\ {\rm  or}\ \sin^2[\frac{k}{\hbar}(t-t_R)],
\nonumber\\
\tau_N &\simeq& \tau_{\rm sync}
\equiv \frac{\hbar}{2k}\times \cos^{-1}(1-2P_0),
\label{sync}
\end{eqnarray}
as the MFS (\ref{mf}) predicts.
In fact, Fig.\ref{fig4} shows that $\tau_N$
saturates to $\tau_{\rm sync}$ for
$N \raisebox{-0.5ex}{$\stackrel{>}{\sim}$} V/2$.
Thus $k=0.1 V_0$ is ``large", where the coupling
to neighbors via $\Delta V_{ij}^\gamma$ is negligible.
Here the system is in the single-body regime.

\subsection{Small hopping amplitude $k$}

\begin{figure}
  \begin{picture}(0,150)
    \put(-50,0){\epsfxsize 280pt
    \epsfbox{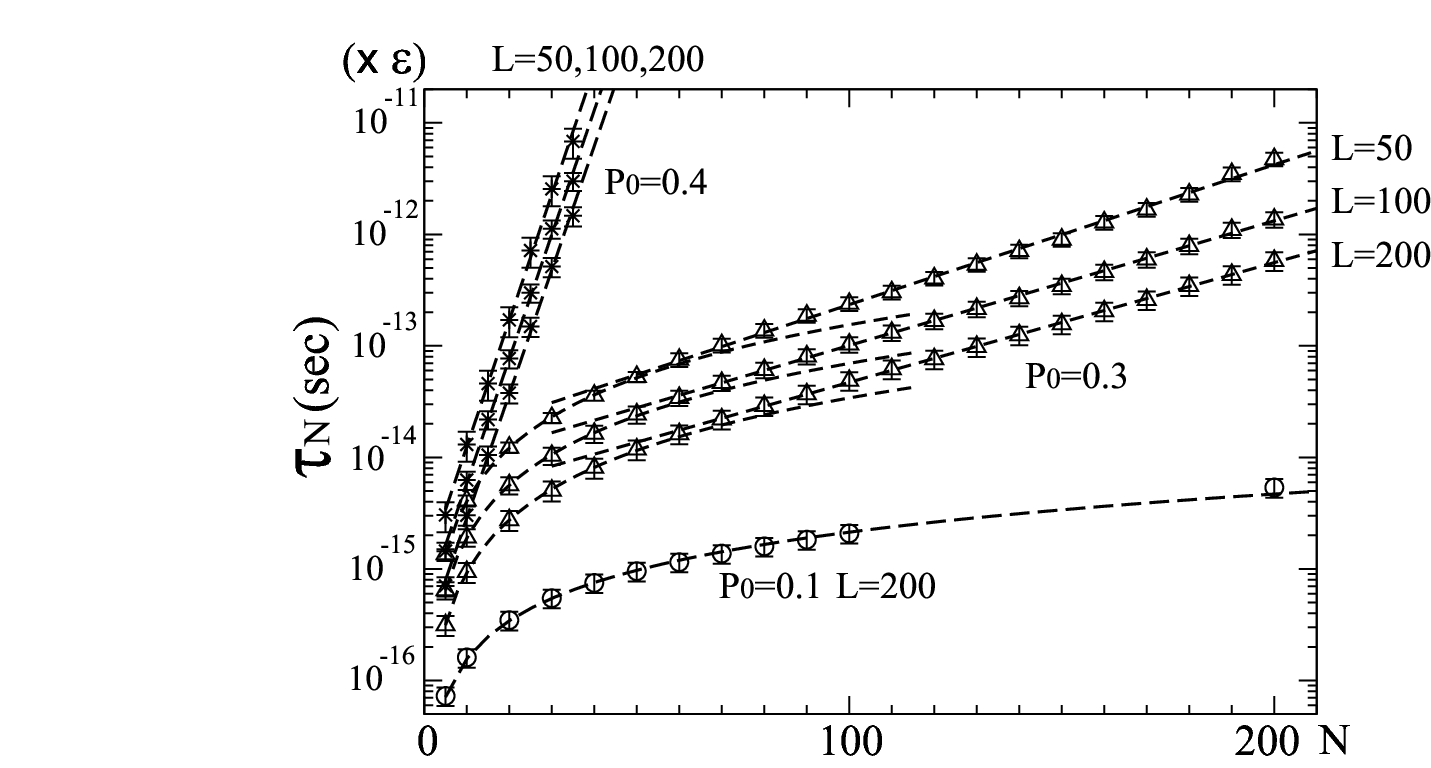}}
  \end{picture}
  \caption{$\tau_N$ vs. $N$ 
for $k=0.01 V_0$ with RIS/LR. 
  $\tau_N$ exhibits a crossover from the power law
  ($\propto N^{b'}, b'\simeq 1.56$ for $P_0=0.3$; dashed curves)
  to the exponential law (\ref{exp}) (dashed straight
   lines).
  }
\label{fig5}
\end{figure}

Let us next consider the case of small $k$'s.
For small $k$'s,
$\Delta V_{ij}^\gamma$ becomes relevant.
In Fig.\ref{fig5} we plot $\tau_N$ vs. $N$ at $k= 0.01 V_0$.
Here each $|C_{i\alpha}(t)|^2$ basically  sweeps
between  0 and 1 with a basic frequency $\sim k/h$,
but is sometimes ``pulled back" due to $\Delta V^\gamma_{ij}$
(See Fig.2b).
The system is in the many-body regime where the potential
term is relevant.
As $N$ increases, $\tau_N$ in Fig.\ref{fig5} exhibits
a crossover from the power law to the exponential law.
We note that, although the  tubulins are not independent
each other here as explained,
Eq.(\ref{percolation})
may describe this crossover phenomenologically
(with an effective $p_c-p$, etc.).
The exponential behavior
of Fig.\ref{fig5} is fitted as



\begin{eqnarray}
\tau_N &\simeq& a' \exp(c'N),\nonumber\\
a' &\simeq& (0.40\sim 1.30)\times
10^{-14}\times \epsilon\ {\rm sec},\nonumber\\
c'&\simeq& 0.025\sim0.029\ {\rm for}\ P_0=0.3,L=50\sim 200.
\label{exp}
\end{eqnarray}

To study the condition for $\tau_N$ to take realistic values,
extremely long runs are required. So let us assume
(\ref{exp}) to hold  for larger $N$'s as an asymptotic
expression, and extrapolate it to
estimate the minimum size $N$ of the cluster.
For $ k= 0.01 V_0,\ P_0=0.3$ and
$a'= 0.85\times 10^{-14}\times \epsilon$\
sec, $c'= 0.027$ we obtain the minimum $N$
for typical values of the dielectric constant $\epsilon$
in Table 2 below.

\vspace{0.3cm}
\begin{center}
\begin{tabular}{|l|c|c|}\hline
&\ $\tau \ge 10^{-2}$\ sec &\ $\tau \ge 10^{-1}$\ sec
\\ \hline
$\ \epsilon =\ \ \ 1$ &\ $N \ge$ 1030  &\ $N \ge$ 1115
\\ \hline
$\ \epsilon =\ \ 10$ & $\ N \ge\ \ $\  945  &\ $N \ge$
1030 \\ \hline
$\ \epsilon =\; 100$ &\  $N \ge\ \ $\  860  &\  $N \ge\ \
$\  945 \\ \hline
\end{tabular}
\end{center}
\vspace{0.3cm}
Table 2. Minimum size of the cluster for
$\tau > 10^{-2}$sec, $10^{-1}$sec and
various $\epsilon$ at $k = 0.01 V_0$ and $P_0= 0.3$
calculated by Eq.(\ref{exp}).

\begin{figure}
  \begin{picture}(0,150)
    \put(-20,0){\epsfxsize 250pt
    \epsfbox{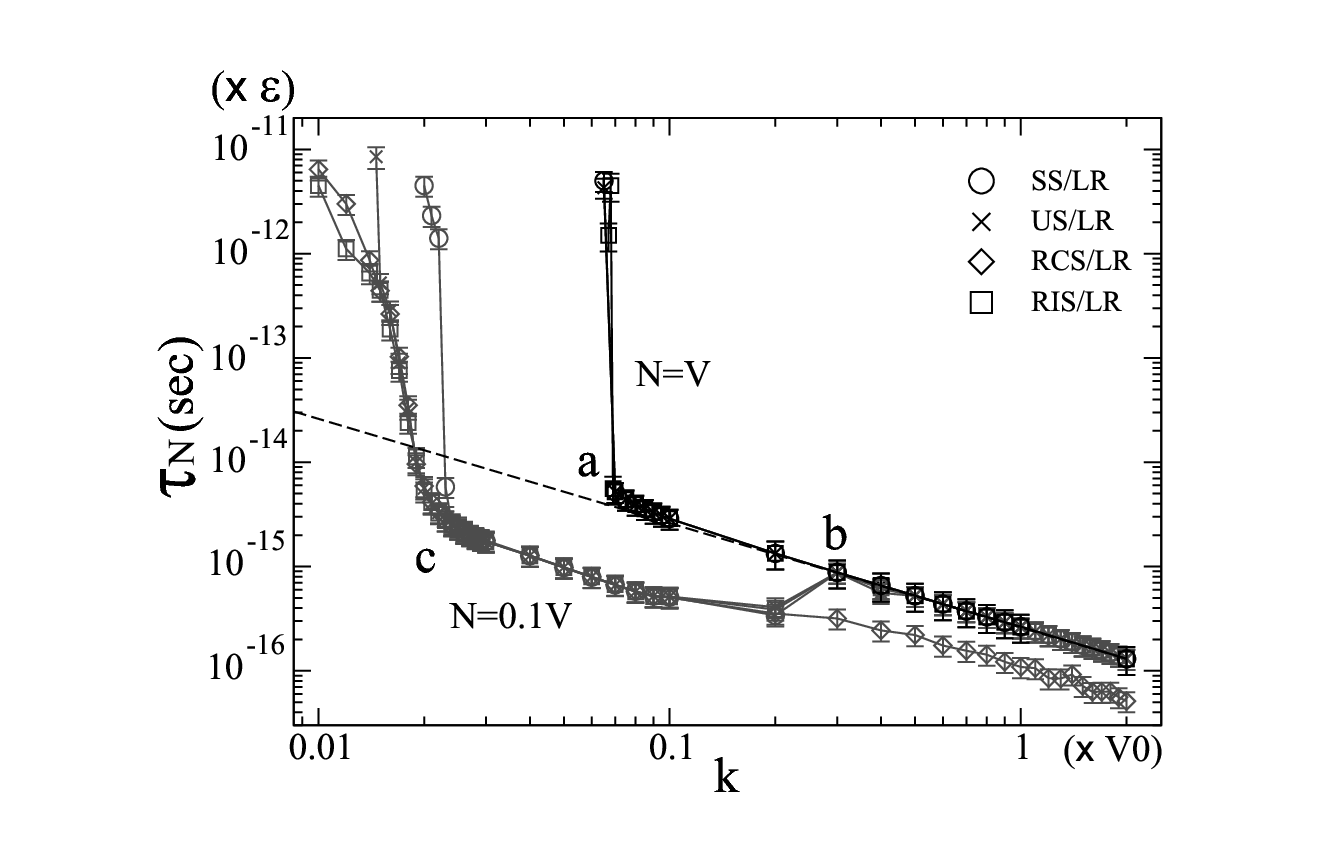}}
  \end{picture}
  \caption{ $\tau_N$ vs. $k$ for $L=200$ and $P_0=0.3$.
  Upper data in black are for $N=V$ and lower data in gray
  are for   $N=0.1V$.
  Dashed  line is $\tau_{\rm sync}$ of (\ref{sync}).
}
\label{fig6}
\end{figure}

\subsection{$k$-dependence and the estimation of $k$}

Let us study the $k$-dependence of $\tau_N$ in detail.
In Fig.\ref{fig6} we plot $\tau_N$ vs. $k$ for
$N=0.1V$ and $N=V$ with $L=200, P_0=0.3$ and LR.
Let us first consider $N=V$.
Simulations with RCS show no reductions
throughout runs due to the random initial configurations.
For the other three initial conditions,
the data (in black) at large $k$'s lie on the line $\tau_N =
\tau_{\rm sync}$  (\ref{sync})
  of the synchronized behavior
  of single cluster explained for Fig.\ref{fig4}.
As $k$ decreases,
they deviate from $\tau_{\rm sync}$
suddenly at the point marked as $a$ with $k=k_a \simeq 0.07 V_0$
As $k \rightarrow 0$, $\tau_N$ blows  up.
This is because these
SS, US, and RIS become exact eigenstates of  $H$ at $k=0$;
each initial eigenstate has time-independent
$|C_{i\alpha}(t)|^2 (= 0,1)$, hence no reductions
are possible.
Let us next consider $N=0.1V$.
The SS, US, and RIS data start from $\tau_{\rm sync}$
at large $k$ and deviate
from it downward  at the point $b$
and join the RCS data.   They start to blow up at $c$.
The decrease of $\tau_{N}$ at $b$ reflects generations
of plural clusters.

\begin{figure}
  \begin{picture}(0,160)
    \put(0,0){\epsfxsize 220pt
    \epsfbox{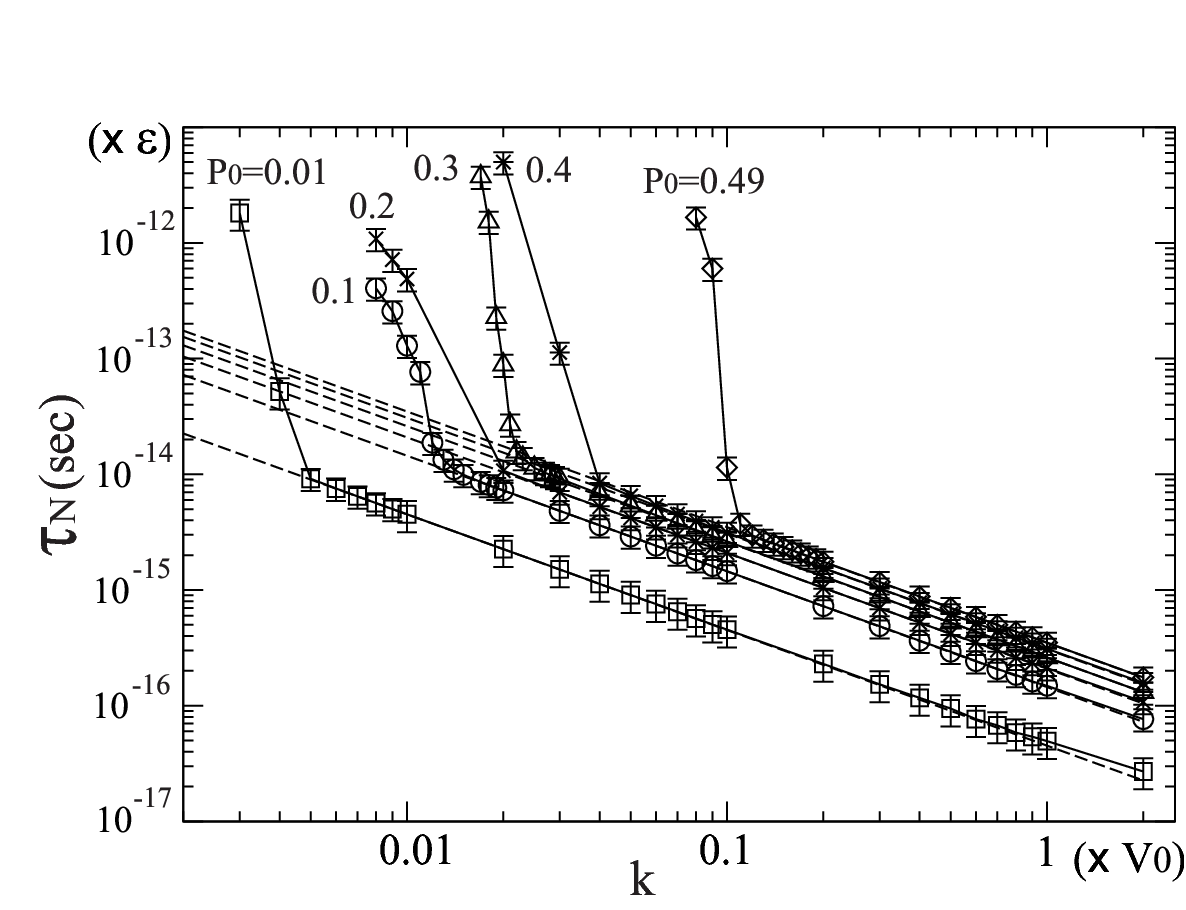}}
  \end{picture}
  \caption{ $\tau_N$ vs. $k$ for various values of $P_0$,
  $L=200, N=V/2$  and  RIS/LR.
Dashed  lines are $\tau_{\rm sync}$ of (\ref{sync}).
}
\label{fig7}
\end{figure}

In Fig.\ref{fig7} we  plot $\tau_N$ vs. $k$
for various values of $P_0$ with RIS/LR, $N=V/2, L=200$.
The data show systematic

\begin{figure}
  \begin{picture}(0,150)
    \put(0,0){\epsfxsize 170pt
    \epsfbox{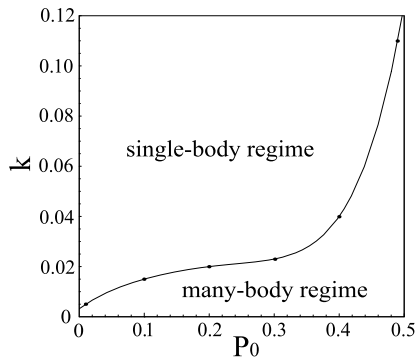}}
  \end{picture}
  \caption{
Two regimes in the $P_0-k$ plane separated by $k=k_a(P_0)$.
The smooth
curve separating two regimes is an interpolation
of $k=k_a(P_0)$ determined from Fig.\ref{fig7} (six small circles).
}
\label{fig8}
\end{figure}

\noindent
dependence on $P_0$.
As $P_0$ increases, the reset zone becomes narrow and
$\tau_N$ at fixed $k$ increases.
The point of blowing up, $k=k_a$, also increases as $P_0$.

In Fig.\ref{fig8} we plot an interpolating
curve of $k_a$ in  Fig.\ref{fig7}
as a function of $P_0$. It separates two regimes:
(i) single-body regime for $k > k_a$ where
the kinetic term in $H$ dominates over the potential term
and the dynamics of each tubulin is well described by
the synchronized motion of Eq.(\ref{sync})
and (ii) many-body regime for $k < k_a$ where
the potential
term dominates over the kinetic term.
As Figs.\ref{fig6},\ref{fig7} show,
the change between these two regimes is very sharp.

Let us estimate the value of  $k$.
For this purpose, we assume
a hydrogen-like wave function for
$|\alpha\rangle$ and $|\beta\rangle$ states;
\begin{eqnarray}
\langle \vec{r}|\gamma\rangle \propto
\exp(-\frac{|\vec{r}-\vec{r}_{\gamma}|}{\ell})
\end{eqnarray}
Then, by a straightforward calculation, one obtains
\begin{eqnarray}
k&\simeq&\frac{\langle \alpha | \frac{e^2}
{|\vec{r}-\vec{r}_{\alpha}|}|\beta \rangle
-\langle \alpha | \beta \rangle \langle \alpha |
\frac{e^2}{|\vec{r}-\vec{r}_{\beta}|}|\alpha \rangle}
{\epsilon\ ( 1-\langle \alpha | \beta \rangle^2)}.
\label{k}
\end{eqnarray}

\begin{figure}
  \begin{picture}(0,140)
    \put(10,0){\epsfxsize 200pt
    \epsfbox{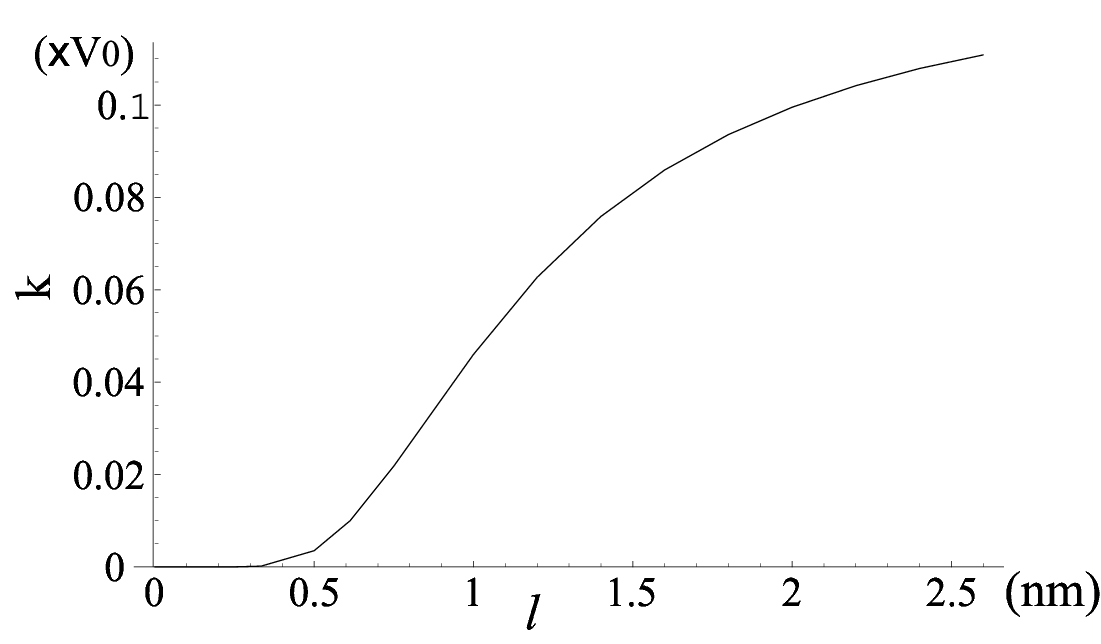}}
  \end{picture}
  \caption{$k$ vs. $\ell$ given by
  Eq.(\ref{k}). Numerically, it gives
$k =0.1 V_0$ for $\ell = 2.0$nm and
$k =0.01 V_0$ for $\ell = 0.61$nm.
  }
\label{fig9}
\end{figure}

In Fig.\ref{fig9} we present $k$ vs. $\ell$ calculated
from Eq.(\ref{k}) by putting 
$|\vec{r}_{\alpha}-\vec{r}_{\beta}|$=4nm as shown
in Fig.\ref{fig1}b. It seems that the reasonable value of
$k$ may be within the range $0.01 V_0\
\raisebox{-0.5ex}{$\stackrel{<}{\sim}$}\ k\
\raisebox{-0.5ex}{$\stackrel{<}{\sim}$}\ 0.1V_0$.

\section{Conclusion and Discussion}
\setcounter{equation}{0}

If we write the time period of a physiologically
reasonable value of each moment in the stream of
consciousness as $\tau_{\rm con}$,
a necessary condition that the scenario of
Hameroff and Penrose\cite{hameroffandpenrose}
becomes successful reads
\begin{eqnarray}
\tau_{\rm{dec}}\ \raisebox{-0.5ex}{$\stackrel{>}{\sim}$}\
\tau_N\ \raisebox{-0.5ex}{$\stackrel{>}{\sim}$}\
\tau_{\rm con}.
\label{scenario}
\end{eqnarray}
For $\tau_{\rm con}$ it is reasonable to
estimate as $\tau_{\rm con}
\simeq 10^{-1} \sim 10^{-2}$ sec.
Then the exponential law (\ref{exp}) for $\tau_N$
at small $k$
shows that, in  order to satisfy the second inequality
of (\ref{scenario}), one needs
$N$ \raisebox{-0.5ex} {$\stackrel{>}{\sim}$} $1000$
as Table 2 shows.
This cluster size is realistic enough, because
it is the same order or smaller than
the typical size of microtubule,
$V = (100 \sim 1000) \times 13$, explained in Sect.1.
Thus, our present analysis shows that the second
inequality holds.
Concerning to the first inequality of (\ref{scenario}),
estimation of $\tau_{\rm dec}$ is crucial, but, as stated
in Sect.1,  it is still controversial\cite{tegmark,hagan}.
We need reliable estimation of $\tau_{\rm dec}$.
Here we point out that
a possible quantum ordered state of a microtubule or a set
of microtubules with spontaneous symmetry breaking and
an off-diagonal long-range order may give rise to
an extra stability of coherence against
decoherent fluctuations due to the environment,
leading to much larger values of $\tau_{\rm dec}$.
This possibility is pointed out by Penrose\cite{penrose} and
others in a general point of view.
Because there is now an explicit quantum model (\ref{hamiltonian}),
one may explore such possibility in a quantitative manner
using  conventional techniques of quantum field theory and
statistical mechanics.
This is certainly an interesting future problem.

In conclusion, we set up a quantum field theory of microtubule
and estimated the self-reduction time $\tau_N$.
$\tau_N$ exhibits systematic behaviors;
the   power law in the single-body regime for $k > k_a$,
or  the  exponential law in the many-body regime for $k < k_a$.
$\tau_N$ can take arbitrary
large values if one imposes unrealistic conditions like $k \rightarrow 0$,
$N \rightarrow \infty$, or $P_0 \rightarrow 0.5$.
Further experimental studies
of microtubules, e.g., determination  of $k$,
may lead us to better understanding of their behaviors;
We may then be ready  to answer some interesting  questions like
whether they are
sitting on ``the edge of chaos" discussed
in the study of complex systems\cite{kauffman}.

Concerning to the approximation(variational method),
we point out a possibility that
the  approximate solution plus self-reductions as we imposed
may mimic the exact solutions of Schr\"odinger equation.
Then we do not need extra coupling to quantum gravity claimed 
by Penrose.
It is interesting to investigate whether
the conventional quantum theory itself
exhibits (quasi)reductions
by letting its subsystems make self-(quasi)reductions\cite{qe}.


We thank
Eiichi Shimizu, Fukutaro Sakai, Yoshitake Kuruma, Hisashi Itoh,
Hiroyuki Kawasaki, Tsutomu Yamada, Yuki Nakano, and Takashi
Odagaki for discussions and correspondences.




\begin{references}


\bibitem{hopfield}
J.J. Hopfield, Proc.Natl.Acad.Sci.USA {\bf 79}, 2554 (1982).
For other neural networks, see, e.g.,
S. Haykin,``Neural Networks; A Comprehensive Foundation",
Macmillan Pub.Co., 1994.

\bibitem{hameroffandpenrose}
S.R. Hameroff and R. Penrose, J. Consciousness Studies {\bf 3}, 36 (1996).

\bibitem{tuszynski}
J.A. Tuszynski, J.A. Brown, and P. Hawrylak,
Phil. Trans. R. Soc. Lond. {\bf A 356}, 1897 (1998).



\bibitem{rasmussen} S. Rasmussen, H. Karampurwala, R. Vaidyanath,
K.S. Jensen, and S. Hameroff, Physica {\bf D42}, 428 (1990).


\bibitem{penrose} R. Penrose, ``Shadows of the mind"
(Oxford Univ. Press, 1994).



\bibitem{tegmark} M. Tegmark, Phys. Rev. {\bf E61}, 4194 (2000).

\bibitem{hagan}
S. Hagan, S.R. Hameroff and J.A. Tuszy\'nski, Phys. Rev. {\bf E65},
061901 (2001).


\bibitem{drm}
A. Bassi and G. Ghirardi, Phys. Rep.{\bf 379}, 257 (2003), 
and references cited therein.

\bibitem{qe}
For a similar proposal of internal quantum measurements
for each living cell, see
J. McFadden,``Quantum Evolution"(Harper Collins, 2000), Chap.11.

\bibitem{wannier}
This is in strong contrast to the related classical
antiferromagnetic Ising model on a
two-dimensional triangular lattice, which has the huge degeneracy
$2^{0.3 N}$ calculated in
G.H. Wannier, Phys. Rev. {\bf 79}, 357 (1950).

\bibitem{cluster}
J. Hoshen and  R. Kopelman,  Phys. Rev. {\bf B14}, 3438 (1976);
K. Binder and D.W. Heermann,
``Monte Carlo Simulation in Statistical Physics (3-ed)"
(Springer-Verlag, 1997).

\bibitem{percolation}
D. Stauffer and A. Aharony, ``Introduction to percolation theory"
(Taylor and Francis, 1994).

\bibitem{mershin}
A. Mershin, A.A. Kolomenski, H.A. Schuessler
and D.N. Nanopoulos, Biosystems {\bf 77}, 73 (2004).

\bibitem{minoura}
I. Minoura and E. Muto, Biophys. J. {\bf 90}, 3739 (2006).

\bibitem{kauffman}
S. Kauffman, ``At Home in the Universe: The Search for the Laws of
Self-Organization and Complexity" (Oxford Univ. Press, 1995).


\end{references}
\end{document}